\documentclass[final,1p,times]{elsarticle} % do not change
\usepackage{graphicx} % do not change
\usepackage{amssymb} % do not change
\usepackage{amsthm} % do not change
\usepackage{lineno} % do not change

\journal{Nuclear Physics A} % do not change
\begin{document} % do not change

\begin{frontmatter} % do not change

\title{The Transition Radiation Detector for ALICE at LHC}
\author{MinJung Kweon$^{a}$ for the ALICE TRD Collaboration}

\address[a]{Physikalisches Institut, Universit\"{a}t Heidelberg, % label [a]
Heidelberg, 69120, Germany}

\begin{abstract} % do not change
The Transition Radiation Detector (TRD) for the ALICE experiment at the Large Hadron Collider (LHC) identifies electrons in p+p and in the challenging high multiplicity environment of heavy-ion collisions and provides fast online tracking for the ALICE Level1 trigger. The TRD is designed to have excellent position resolution and pion rejection capability. Presently, six of the 18 TRD supermodules are installed in the ALICE central barrel. In 2008, four supermodules were installed and commissioning of the detector using cosmic ray tracks was successfully performed. We briefly describe the design of the detector and report on the performance and current understanding of the detector based on these data.   
\end{abstract} % do not change

\end{frontmatter} % do not change

%% QM09: we keep linenumbers at least for initial version
%\linenumbers % do not change

\section{Introduction}

ALICE (A Large Ion Collider Experiment) is a general-purpose heavy-ion experiment designed to study the physics of strongly interacting matter and the quark-gluon plasma in nucleus-nucleus collisions at the LHC \cite{alice_ppr_vol2}. It will study the global properties with hadron production and correlations, and probe the properties of the medium with the collision products such as heavy quarkonia, open charm and beauty, light vector mesons, thermal leptons, direct-$\gamma$, jets and high-$p_{T}$ hadrons. One of the powerful probes of the created QCD medium is heavy quarkonia, whose suppression or enhancement \cite{SHM_model} is sensitive to the screening of color charge due to deconfinement and to statistical recombination. Thus measurements of leptons from their decay are crucial.

An important task of the Transition Radiation Detector (TRD) for the ALICE experiment is to supplement the Time Projection Chamber (TPC) electron/pion identification by a pion rejection factor of the order of 100 at momenta in excess of 1 GeV/c. In addition, by measurement of energy loss, the TRD improves the identification of other charged particles \cite{alice_ppr_vol2}. The TRD also provides space points for the global central barrel tracking together with the Inner Tracking System (ITS) and the TPC. Due to its large lever arm it improves the overall momentum resolution, especially at high momentum, where the resolution is expected to be $\sim$3.5\% at 100 GeV/c \cite{alice_ppr_vol2}. Additionally, the TRD provides a fast trigger for single/pairs of electrons and cluster of high-$p_{T}$ tracks, thus it allows us to study rare probes such as high-$p_{T}$ $\textit{J}$/$\psi$, $\Upsilon$ and high-$E_{T}$ jets.

\section{Working Principle and Design}

\begin{figure}[ht]
\begin{minipage}[b]{0.5\linewidth}
\centering
\includegraphics[scale=0.9]{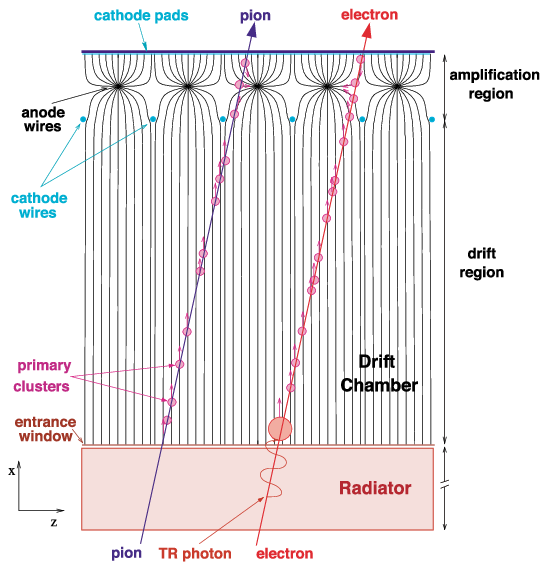}
\caption{Cross-sectional ($r-z$ plane) view of one TRD chamber showing the passage of a pion/electron.} 
\label{design_principle}
\end{minipage}
\hfill
\begin{minipage}[b]{0.47\linewidth}
\centering
\includegraphics[scale=0.8]{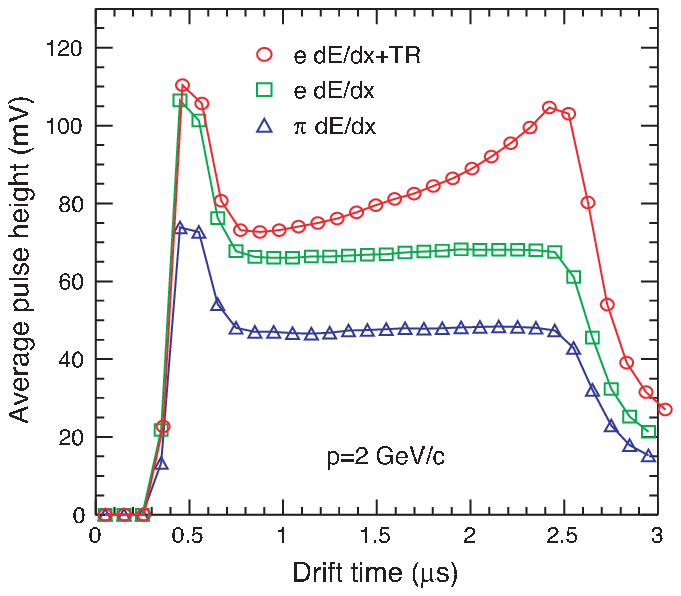}
\caption{Average pulse height as a function of drift time for pions and electrons w/wo radiator \cite{avg_pulseheight}.}
\label{pulse_height_prototype}
\end{minipage}
\end{figure}

The required pion rejection of a factor 100 is obtained by production of transition radiation (TR) which is only generated by electrons within the relevant momentum range. The relativistic electron ($\gamma$$\ge$1000) radiates photons in the X-ray range when it traverses a boundary between media of different refraction indices. These TR photons ($\le$ 30 keV) are absorbed by the high-$Z$ gas mixture (Xe-based) of high photo-absorption cross section. The Fig. \ref{design_principle} illustrates the working principle and design of the TRD chamber. Considering TR photon detection and tracking, the chamber consists of polypropylene fibres/form sandwich radiator (48 mm), drift (30 mm) and multi-wire proportional (7 mm) section with cathode readout pad. On top of the signal due to the ionization of the gas by traversing charged particle, the characteristic peak is produced by the TR photons at large drift time for electron as it is shown in Fig. \ref{pulse_height_prototype}. This allows us to separate electrons from pions. The 540 chambers of the TRD are arranged in 18 supermodules, which surrounds the TPC in the central barrel of ALICE, containing 5 stacks along the longitudinal and 6 layers along the radial direction. It has 694 m$^{2}$ active area and 28 m$^{2}$ gas volume of Xe/CO$_{2}$ (85/15). The total radiation length is $\sim$24\% of $X_{0}$ and the total weight is $\sim$30 tons.

The TRD front-end electronics (FEE) is directly mounted on the back panel of the chamber. Groups of 18 pads are connected by short cables to a Multi-Chip Module (MCM) which comprises the Pre-Amplifier and Shaper Amplifier (PASA) and the Tracklet Processor (TRAP) \cite{trap}. The PASA has a 120 ns shaping time, a gain of 12.4 mV/fC, and an equivalent noise charge of 850 electrons at 25 pF input capacitance. The TRAP chip comprises 21 channels each with a 10 MHz 10 bits ADC, four stages of digital filters, event buffers, readout interface and local tracking unit (Preprocessor and four 120 MHz CPUs) which allows to calculate the inclination of a track in the bending direction as well as to measure the total charge deposited along the track. One Read-Out Board (ROB) consists of 17 or 18 MCMs and 6 or 8 ROBs are mounted on each chamber. Each chamber has one Linux based Detector Control System (DCS) board which controls FEE and two optical readout interface modules (ORI) for data shipping.

The local track segments (tracklets) and the raw data acquired by the TRAP are sent to the global tracking unit (GTU) via 2.5 Gbps ORI. Based on the tracklets collected, the GTU performs transverse momentum reconstruction and electron identification. After finding high-$p_{T}$ tracks and identifying electrons, various trigger schemes are applied for di-electron decays and jets. Then this trigger contribution is sent to the ALICE central trigger processor (CTP) within 6.1 $\mu s$ to drive the Level-1 trigger decision. Such fast processing is possible due to the massive parallel hardware architecture of the GTU whose core is FPGA-based. It is capable of processing up to 20k tracklets within 2 $\mu s$ (maximum 16k tracklets for $dN_{ch}/dy$ = 8000 events with a $p_{T}$ threshold of 2.3 GeV/c). The GTU also forwards raw data to the DAQ when Level-2 is accepted.

\section{Integration, Installation and Commissioning}

Supermodule integration is done layer-wise with 30 chambers and each layer undergoes tests of electronics, gas tightness, high-voltage distribution and cooling. It is concluded by several days of cosmics data taking to produce a calibration data set. The first supermodule was installed in October 2006. Noise measurements after installation show that we achieved an average noise level of 1.1 ADC counts, which is close to the design goal. The fraction of dead channels is less than 0.1\%.

In total, four supermodules were installed in ALICE in 2008 and participated in the cosmic ray data taking together with other detectors. Besides the proof of combined operation, these runs are used to gather reference data for alignment and calibration from traversing cosmic particles. The cosmic trigger decision was based on coincident hits in the Time Of Flight (TOF) detector and GTU Level-1 trigger. For the first time we used the full chain of trigger sequence for the TRD within the ALICE setup and the GTU Level-1 trigger was the first running Level-1 trigger in ALICE. Level-1 rejection from Level-0 (originated by TOF coincident hits) was a factor 20 and the Level-1 rate was $\sim$0.05 Hz with a purity better than 85\%. In total 55k tracks were recorded in the TRD under tight constraints on the cosmic ray topology, requiring tracks close to horizontal at 60 m below the earth surface. Before the LHC injection test in 2008, we were ready for data taking. 

\begin{figure}[ht]
\begin{minipage}[b]{0.5\linewidth}
\centering
\includegraphics[scale=1]{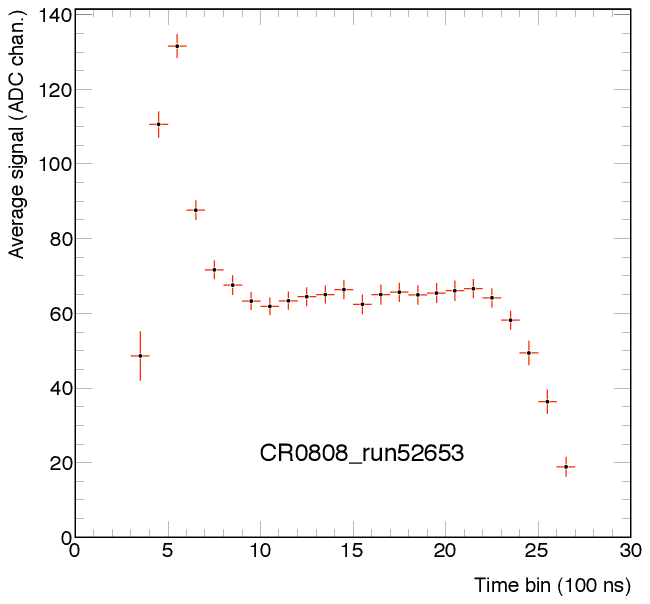}
\caption{Average pulse height as a function of time}
\label{pulse_height}
\end{minipage}
\hspace{0.5cm}
\begin{minipage}[b]{0.45\linewidth}
\centering
\includegraphics[scale=0.8]{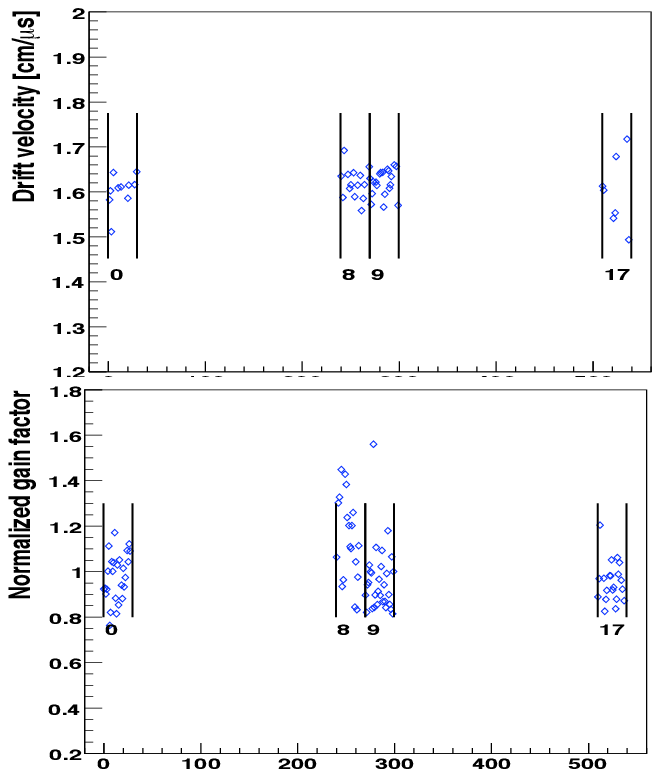}
\caption{Drift velocity and normalized gain factor as a function of detector number}
\label{trd_gain_driftvelocity}
\end{minipage}
\end{figure}

The tracking and PID algorithms of the TRD rely on the knowledge of several calibration constants which depend on environment temperature and gas pressure, gas composition and chamber geometry. The calibration constants are the drift velocity of the electrons, the time-offset of the signal and the gas gain. During data taking, a first calibration is performed as an online procedure and later also done offline. The drift velocity and time offset are determined with the average signal as a function of time (Fig. \ref{pulse_height}). With the assumption that the cosmic rays are uniformly distributed over the detector, a gas gain variation of about 16\% was found over the chambers. Fig. \ref{trd_gain_driftvelocity} shows the gain factors for the four supermodules. The results obtained online were consistent with those obtained offline (Fig. \ref{trd_gain_driftvelocity}) after the tracking in a second pass calibration. The extracted drift velocity values has a variation of 3.3\% over the chambers. 

The obtained spatial resolution within the TRD chambers is $\approx$350 $\mu m$ at 0$^{\circ}$, which is close to the design goal.

\begin{figure}[ht]
\begin{minipage}[b]{0.45\linewidth}
\centering
\includegraphics[scale=0.66]{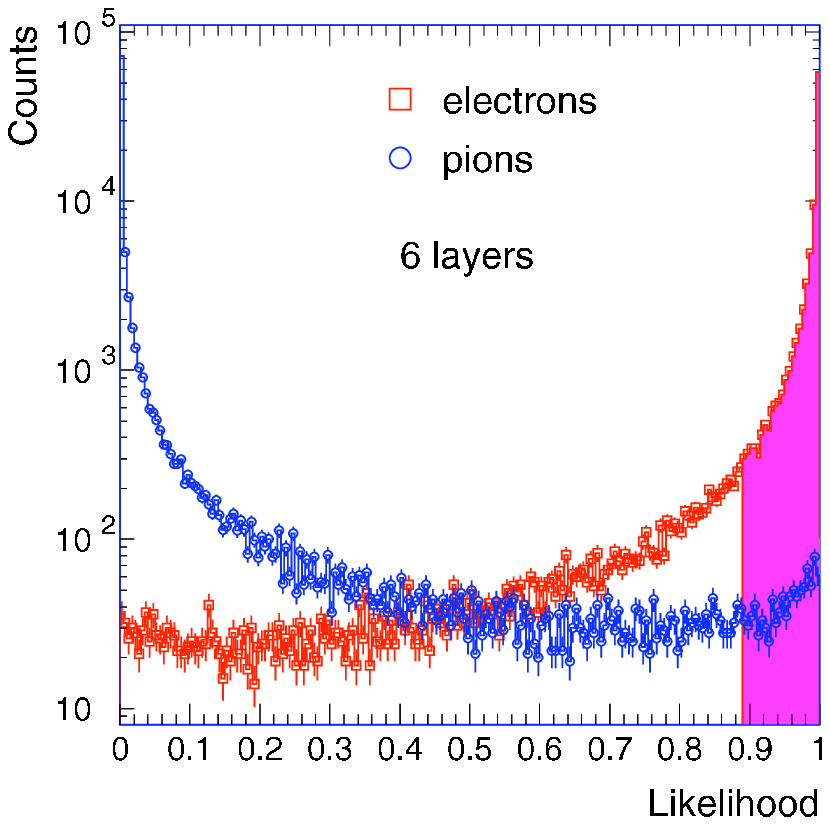}
\caption{Distributions of the likelihood for electrons and pions with a momentum of 2 GeV/c, obtained from the total energy deposit.}
\label{liklihood}
\end{minipage}
\hspace{0.5cm}
\begin{minipage}[b]{0.50\linewidth}
\centering
\includegraphics[scale=0.75]{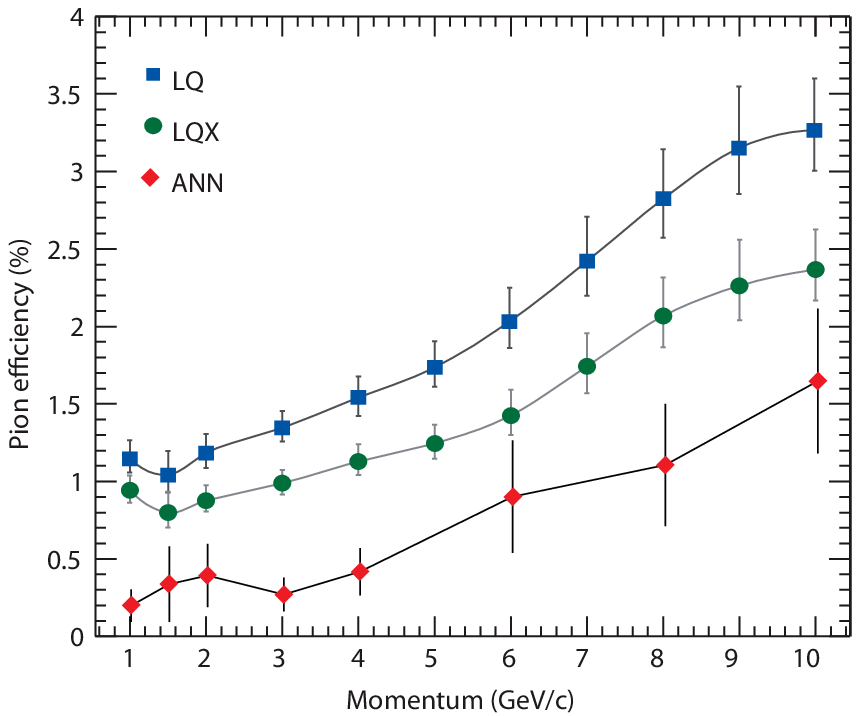}
\caption{Measured pion efficiency as a function of momentum for three methods: likelihood on total charge, bidimensional likelihood \cite{pion_eff} and neural networks \cite{Alex_private_communication}.}
\label{pion_efficiency}
\end{minipage}
\end{figure}

As a part of commissioning, in 2004 and 2007 there were test beam measurement at CERN PS with electron and pion beams. The likelihood distributions for six layers, based on the total energy deposit in one layer are shown in Fig. \ref{liklihood} for the momentum of 2 GeV/c. Cuts of given electron efficiency are imposed on the likelihood value and the pion efficiency $\pi_{eff}$ is calculated. The momentum dependence of the pion efficiency calculated with different likelihood methods and for the neural networks is shown in Fig. \ref{pion_efficiency}. It demonstrates that we exceed the design goal of factor 100 pion rejection for isolated tracks for momenta less than 10 GeV/c.

Seven out of the total 18 supermodules of the TRD will be ready in the ALICE setup when cosmic ray data taking resumes in July 2009 and will contribute to physics results with beams expected to start in fall 2009. The completion of the TRD setup with all 18 supermodules installed is planned during the next long shutdown of the LHC.

%% end of main text

\section*{Acknowledgments} 
We acknowledge the ALICE collaboration, the installation and support team at CERN. This work has been supported by German BNBF.

 % do not change 

\begin{thebibliography}{00} % do not change 
   
\bibitem{alice_ppr_vol2} Alice Physics Performance Report VolumeII, ALICE Collaboration, {\it J. Phys. G: Nucl. Part. Phys.} {\bf 32} (2006)
1295-2040
 
\bibitem{SHM_model} A. Andronic, P. Braun-Munzinger, K. Redlich, J. Stachel, {\it J. Phys. G: Nucl. Part. Phys.} {\bf 35} (2008) 104155

\bibitem{avg_pulseheight} A. Andronic, {\it NIM A} {\bf 522} (2004) 40-44
\bibitem{trap} V. Angelov et al (ALICE TRD Collaboration), {\it NIM A} {\bf 563} (2006) 317-320

\bibitem{pion_eff} R. Bailhache, C. Lippmann (ALICE TRD Collaboration), {\it NIM A} {\bf 563} (2006) 310-313
\bibitem{Alex_private_communication} private communication with Alexander Wilk 
\end{thebibliography}
\end{document}